\documentclass[11pt]{article}
\usepackage[latin10]{inputenc}
\usepackage{amsmath}
\usepackage{amssymb}
\usepackage{graphicx}
\usepackage{esint}

\makeatletter

\usepackage{amsfonts}
\providecommand{\U}[1]{\protect\rule{.1in}{.1in}}

\makeatother

\begin{document}
\begin{titlepage} \vspace{0.3cm}
 \vspace{1cm}
 
\begin{center}
\textsc{\Large{}{}{}{}{}{}{}{}{}{}\ \\[0pt] \vspace{0mm}
 Mimetic Inflation }{\Large{}{}and Self-reproduction{} }\\[0pt]
\textsc{\Large{}{}{}{}{}{}{}{}{}{}\ \\[0.0pt] }{\Large\par}
\par\end{center}

\begin{center}
 
\par\end{center}


\begin{center}
\vspace{35pt}
 \textsc{ A. H. Chamseddine$^{~a,b}$, M. Khaldieh$^{~a}$, V. Mukhanov$^{~a}$}\\[15pt] 
\par\end{center}

\begin{center}
{$^{a}$ \textit{Ludwig Maxmillian University, \\[0pt] Theresienstr.
37, 80333 Munich, Germany}}\\[0pt] {\small{}{}{}{}{}{}{}{}{}} 
\par\end{center}


\begin{center}
{$^{b}$ {\small{}{}{}{}{}{}{}{}{}{}}\textit{Physics Department,
}\\
 \textit{ American University of Beirut, Lebanon}}\\
 
\par\end{center}

\vspace{1cm}
 
\begin{abstract}
It is shown how self-reproduction can be easily avoided in the inflationary universe, even when inflation starts at Planck scales. This is achieved by a simple coupling of the inflaton potential with a mimetic field. In this case, the problem of fine-tuning of the initial conditions does not arise, while eternal inflation and the multiverse with all their widely discussed problems are avoided. 
\end{abstract}
\end{titlepage}

\section{Introduction}

One of the most important phenomena that accompanies slow-roll inflation
based on scalar field potentials is the regime of self-reproduction.
If we want to begin inflation at Planck scale to avoid fine-tuning
of the initial conditions, this regime seems to be unavoidable. This,
in turn, leads to an unwarranted situation where most of the space
is still in the regime of accelerated expansion at all times, resulting
in a picture of the multiverse, with a possible loss of predictability
of the theory.

If we consider a homogeneous scalar field $\varphi$ with potential
$V\left(\varphi\right)$ in a flat expanding universe with metric
\begin{equation}
ds^{2}=dt^{2}-a^{2}\left(t\right)\delta_{ik}dx^{i}dx^{k},\label{eq:1a}
\end{equation}
then the equations for this field and the scale factor $a\left(t\right)$
are 
\begin{align}
 & \ddot{\varphi}+3H\dot{\varphi}+V^{\prime}=0,\nonumber \\
 & H^{2}=\frac{1}{3}\left(\frac{1}{2}\dot{\varphi}^{2}+V\right),\label{eq:2a}
\end{align}
where the dot denotes the time derivative with respect to physical
time $t$ and the prime is the derivative of the potential $V$ with
respect to $\varphi$. The Hubble constant is defined as $H\equiv\dot{a}/a$,
we use Planck units here and in the following and we set $8\pi G=1.$
The above equations can be simplified if the potential $V$ satisfies
the slow-roll conditions 
\begin{equation}
\left(\frac{V^{\prime}}{V}\right)^{2}\ll1,\qquad\frac{V^{\prime\prime}}{V}\ll1,\label{eq:3aa}
\end{equation}
In this case we can neglect the second derivative in the first equation
and the kinetic energy in the second equation in (\ref{eq:2a}), so
that these equations become 
\begin{align}
 & 3H\dot{\varphi}+V^{\prime}\approx0,\nonumber \\
 & H^{2}\approx\frac{1}{3}V.\label{eq:4a}
\end{align}
During the cosmological time $t_{H}\simeq H^{-1}$ the classical scalar
field decreases its value by the amount 
\begin{equation}
\delta\varphi_{cl}\simeq\dot{\varphi}t_{H}\simeq\frac{\dot{\varphi}}{H}\sim-\frac{V^{\prime}}{V}.\label{eq:5}
\end{equation}
At the same time, the typical quantum fluctuations at Hubble scale
have amplitude $\pm H$ and if this amplitude is positive, the total
change of the scalar field at scale $H^{-1}$ can exceed $\delta\varphi_{cl}$,
and instead of decreasing, the resulting scalar field can increase
by a value 
\begin{equation}
\Delta\varphi\simeq H-\frac{V^{\prime}}{V}>0.\label{eq:6}
\end{equation}
As a result, there is always an exponentially large number of regions
where the acceleration continues forever, the inflation becomes eternal,
the volume filled by the inflating regions reproduces exponentially
fast. This volume is obviously dominant compared to the volume of
the regions where the inflation ends and a Friedmann universe emerges
thereof, filled by a normal matter\cite{Linde}. This is true for
almost every measure used to compare the volume of inflationary regions
with that of the regions where inflation has ended (see for e.g. \cite{Winitzkii}).
Consequently, we are faced by either the problem of fine-tuning in
its weak form, i.e., require inflation to start below the Planck scale
if we are to avoid self-reproduction, or the problem of eternal inflationary
universe, or even a landscape, thus damaging the predictive power
of inflation \cite{Eternal}. Clearly, we must start inflation at
the Planck scale if we are to avoid any degree of fine-tuning of initial
conditions, because only in this case we do not obtain a small-value
for the probability of inflation from the available parameters all
of which are of order one, regardless of any measure characterizing
the ``probability of initial conditions''\ (whatever that may mean).
The natural question to ask, therefore, is how can we avoid in this
case, self-reproduction of the universe. As follows from (\ref{eq:6}),
the condition for self-reproduction at the Planck scale where $H\sim1$,
can only be violated if the slow-roll condition $V^{\prime}/V\ll1$
is at least, weakly violated as well, i.e. $V^{\prime}/V\sim O(1)$
and is only later restored. This leads to a rather restricted class
of unique potentials for which we can start inflation near the Planck
scale and avoid self-reproducing eternal universe \cite{Mukhanov1}.
In this case, the number of universes generated by inflation can be
huge (or even infinite), but all of them will have the same statistical
properties, so that the predictability of the theory is fully restored,
leaving aside the ``philosophical question''\ about the number
of similar non-observable worlds.

In this paper we will show how the problem of self-reproduction for
arbitrary potentials in models where inflation starts at Planck scales
can be solved by simply coupling the inflaton potential to the mimetic
field. Note that the introduction of the mimetic field in no way implies
that we are considering models with multiple scalar fields, but implies
instead a minimal modification of Einstein gravity. This in turn opens
up a playground to build a new class of mimetic inflation models,
which still belong to the class of ``simple inflation''.

\section{Coupled mimetic and inflaton fields}

The mimetic field $\phi$ introduced in \cite{Mimetic} satisfies
the first order differential equation 
\begin{equation}
g^{\mu\nu}\phi_{,\mu}\phi_{,\nu}=1,\label{eq:7}
\end{equation}
and together with the longitudinal mode of gravity reproduces the
dark matter component in the universe. The solution of this equation
in the synchronous coordinate system: 
\begin{equation}
ds^{2}=dt^{2}-\gamma_{ik}dx^{i}dx^{k},\label{eq:8}
\end{equation}
is very simple and given by $\phi=t$ and thus $\phi$ can be considered
as the time coordinate defining the partitioning of the four-dimensional
manifold by three- dimensional space-like hypersurfaces. In this coordinate
system the invariant 
\begin{equation}
\kappa\equiv\Box\phi=g^{\mu\nu}\phi_{;\mu\nu}\label{eq:9}
\end{equation}
has a simple physical interpretation and is equal to the trace of
the extrinsic curvature of the hypersurfaces with constant $\phi$:
\begin{equation}
\kappa=\frac{1}{2}\frac{\partial}{\partial t}\ln\left(\det\gamma_{ik}\right).\label{eq:10}
\end{equation}
Thus, using term $\kappa\equiv\Box\phi$ in the Lagrangian we modify
gravity without adding an additional scalar degree of freedom. In
previous works, we have already used this invariant to resolve singularities
in the universe and black holes \cite{MimeticSing}. In this work,
we will use it to build simple inflationary scenarios without eternal
inflation.

Consider the theory with action 
\begin{equation}
S=\int d^{4}x\sqrt{-g}\left[-\frac{1}{2}R+\lambda(g^{\mu\nu}\partial_{\mu}\phi\partial_{\nu}\phi-1)+\frac{1}{2}g^{\mu\nu}\partial_{\mu}\varphi\partial_{\nu}\varphi-C\left(\kappa\right)V(\varphi)\right],\label{eq:11}
\end{equation}
where we set $8\pi G=1$ and $\varphi$ is the inflaton field with
potential $V\left(\varphi\right)$ coupled to a function $C\left(\kappa\right)$
which depends only on the d'Alembertian of the mimetic field. The
variation of this action with respect to the Lagrange multiplier $\lambda$
leads to the constraint condition (\ref{eq:7}) for the mimetic field.
The modified Einstein equations obtained by variation with respect
to the metric are 
\begin{equation}
G_{\mu\nu}=R_{\mu\nu}-\frac{1}{2}g_{\mu\nu}R=T_{\mu\nu},\label{eq:12}
\end{equation}
where 
\begin{align}
 & T_{\mu\nu}=2\lambda\partial_{\mu}\phi\partial_{\nu}\phi+g_{\mu\nu}\left(\left(C-\kappa C^{\prime}\right)V-g^{\rho\sigma}\partial_{\rho}\left(C^{\prime}V\right)\partial_{\sigma}\phi\right)\nonumber \\
 & +\left(\partial_{\mu}\left(C^{\prime}V\right)\partial_{\nu}\phi+\partial_{\nu}\left(C^{\prime}V\right)\partial_{\mu}\phi\right)+\partial_{\mu}\varphi\partial_{\nu}\varphi-\frac{1}{2}g_{\mu\nu}\left(g^{\rho\sigma}\partial_{\rho}\varphi\partial_{\sigma}\varphi\right)\label{eq:13}
\end{align}
and the prime always denotes the derivative with respect to the argument
of the corresponding function which depends only on one variable,
either $\kappa$ or $\varphi$, i.e., $C^{\prime}\equiv dC/d\kappa$
and $V^{\prime}\equiv dV/d\varphi$. Next, the $\phi$ equation of
motion is 
\begin{equation}
\partial_{\mu}\left[\sqrt{-g}g^{\mu\nu}\left(2\lambda\partial_{\nu}\phi+\partial_{\nu}\left(C^{\prime}V\right)\right)\right]=0,\label{eq:14}
\end{equation}
and finally the inflaton equation obtained by variation with respect
to $\varphi$ is 
\begin{equation}
\frac{1}{\sqrt{-g}}\partial_{\mu}\left(\sqrt{-g}g^{\mu\nu}\partial_{\nu}\varphi\right)+CV^{\prime}=0,\label{eq:15}
\end{equation}
where the first terms is $\square\varphi$.

\section{Background solutions}

Consider a homogeneous flat universe with metric 
\begin{equation}
ds^{2}=dt^{2}-a^{2}\left(t\right)\delta_{ik}dx^{i}dx^{k}.\label{eq:16}
\end{equation}
In this case, $\varphi=\varphi\left(t\right)$ and the solution of the equation (\ref{eq:7}) is $\phi=t$. The $0-0$ Einstein equation
becomes 
\begin{equation}
\frac{1}{3}\kappa^{2}=2\lambda+\left(C-\kappa C^{\prime}\right)V+\left(C^{\prime}V\right)\dot{}+\frac{1}{2}\dot{\varphi}^{2},\label{eq:17}
\end{equation}
where the dot denotes the derivative with respect to time $t$ and,
as it follows from (\ref{eq:10}), 
\begin{equation}
\kappa=3\frac{\dot{a}}{a},\label{eq:18}
\end{equation}
i.e., $\kappa$ is the tripled Hubble constant $H=3\dot{a}/a.$ Equation
(\ref{eq:14}) can be solved explicitly and gives 
\begin{equation}
2\lambda=\frac{B}{a^{3}}-\left(C^{\prime}V\right)\dot{},\label{eq:19}
\end{equation}
where $B$ is the integration constant quantifying ``mimetic dust''\ and
which we set to zero. Then equation (\ref{eq:17}) simplifies to 
\begin{equation}
\frac{1}{3}\kappa^{2}=\left(C-\kappa C^{\prime}\right)V+\frac{1}{2}\dot{\varphi}^{2},\label{eq:20}
\end{equation}
and equation (\ref{eq:15}) becomes 
\begin{equation}
\ddot{\varphi}+\kappa\dot{\varphi}+CV^{\prime}=0.\label{eq:21}
\end{equation}
These two equations are enough to determine the behavior of $a\left(t\right)$
and $\varphi\left(t\right)$. Taking the derivative of equation (\ref{eq:20})
and using (\ref{eq:21}) we find the rate of change of the Hubble
constant 
\begin{equation}
\dot{\kappa}=-\frac{3}{2}\left(\dot{\varphi}^{2}+\left(C^{\prime}V\right)\dot{}\right),\label{eq:22}
\end{equation}
or, equivalently, 
\begin{equation}
\dot{\kappa}=-\frac{3}{2}\frac{\dot{\varphi}\left(\dot{\varphi}+C^{\prime}V^{\prime}\right)}{1+\frac{3}{2}C^{\prime\prime}V}.\label{eq:23}
\end{equation}
In the rest of this section, we consider several interesting examples
and derive the asymptotical solutions for the homogeneous background.

\subsection{Case A}

To demonstrate the idea, we first consider the nonrealistic case,
where, on one hand, we can avoid self-reproduction and eternal inflation,
but on the other hand, as we show in the next section, the ratio of
tensor to scalar perturbations for the observable scales is too large.
Let's take 
\begin{equation}
C\left(\kappa\right)=\frac{\kappa}{\kappa_{0}},\label{eq:24}
\end{equation}
where $\kappa_{0}$ is a free parameter. In this case the equations
(\ref{eq:20}) and (\ref{eq:21}) simplify to 
\begin{align}
 & \kappa^{2}=\frac{3}{2}\dot{\varphi}^{2},\nonumber \\
 & \ddot{\varphi}+\kappa\left(\dot{\varphi}+V^{\prime}/\kappa_{0}\right)=0.\label{eq:25}
\end{align}
Considering that the field $\varphi$ should decrease during the expansion,
i.e., $\dot{\varphi}<0$ for $\kappa>0$, equation (\ref{eq:24})
gives 
\begin{equation}
\kappa=-\sqrt{\frac{3}{2}}\dot{\varphi}.\label{eq:26}
\end{equation}
Substituting this $\kappa$ into (\ref{eq:25}) and taking into account
that $\ddot{\varphi}=\dot{\varphi}\left(d\dot{\varphi}/d\varphi\right)$
this equation reduces to the first order differential equation 
\begin{equation}
\frac{d\dot{\varphi}}{d\varphi}=\sqrt{\frac{3}{2}}\left(\dot{\varphi}+\frac{V^{\prime}\left(\varphi\right)}{\kappa_{0}}\right)\label{eq:27}
\end{equation}
for $\dot{\varphi}$ as a function of $\varphi$, whose explicit solution
is 
\begin{equation}
\dot{\varphi}=\frac{b}{\kappa_{0}}e^{b\varphi}\intop^{\varphi}V^{\prime}\left(\tilde{\varphi}\right)e^{-b\tilde{\varphi}}d\tilde{\varphi},\label{eq:28}
\end{equation}
where, to avoid cluttering of numerical factors, we have set $b=\sqrt{3/2}$.
The lower limit in the integral corresponds to the choice of the integration
constant multiplied by $e^{b\varphi}$, and since $\varphi$
decreases with time, this solution decays. If we neglect this integration
constant and integrate by parts in (\ref{eq:28}), we can rewrite
the attractor solution of (\ref{eq:27}) as 
\begin{equation}
\dot{\varphi}=-\frac{1}{\kappa_{0}}\left(V'+\frac{V^{\prime\prime}}{b}+\frac{V^{\prime\prime\prime}}{b^{2}}+...\right).\label{eq:29}
\end{equation}
In the case of power-law potentials $V\propto\varphi^{n}$, the series
on the right hand side has only a finite number of terms and the equation
can further be integrated to obtain $\varphi\left(t\right)$. Let
us consider the following potential
\begin{equation}
V\left(\varphi\right)=\frac{1}{2}m^{2}\varphi^{2},\label{eq:30}
\end{equation}
and set $\kappa_{0}=m$ to simplify the formulas. The equation (\ref{eq:29})
is then 
\begin{equation}
\dot{\varphi}=-\frac{m}{b}\left(1+b\varphi\right)\label{eq:31}
\end{equation}
and can be easily integrated to obtain 
\begin{equation}
\varphi\left(t\right)=\left(\varphi_{i}+\frac{1}{b}\right)e^{-mt}-\frac{1}{b},\label{eq:32}
\end{equation}
where $\varphi_{i}=\varphi\left(t=0\right)$. If we start at the Planck
scale, at $t=0$, where $\dot{\varphi}^{2}\simeq1$, then $\varphi_{i}\simeq m^{-1}\gg1$
for $m\ll1$. From equations (\ref{eq:20}) and (\ref{eq:23}) it
follows that in this particular case 
\begin{equation}
\frac{\dot{\kappa}}{\kappa^{2}}=-\frac{1}{1+b\varphi}.\label{eq:33}
\end{equation}
Recalling that $\kappa=3H$ and 
\begin{equation}
H^{2}=\varepsilon/3,\quad\dot{H}=-\frac{1}{2}\left(\varepsilon+p\right),\label{eq:34}
\end{equation}
where $\varepsilon$ and $p$ are the energy density and pressure
(see, e.g., \cite{Mukhanovbook}), the expression (\ref{eq:33}) can
be rewritten as 
\begin{equation}
\frac{\varepsilon+p}{\varepsilon}=\frac{2}{1+b\varphi}.\label{eq:35}
\end{equation}
Thus for $\varphi\gg1$ we have $\left(\varepsilon+p\right)/\varepsilon\ll1$,
i.e., $p\approx-\varepsilon$ and the universe undergoes exponential
expansion. When the field $\varphi$ drops below one, inflation ends
and at $\left\vert \varphi\right\vert \ll1$ the universe is dominated
by matter with the ultra-hard equation of state $p\approx+\varepsilon$.
Integrating (\ref{eq:26}) we find that 
\begin{equation}
a=a_{i}\exp\left(\frac{1}{\sqrt{6}}\left(\varphi_{i}-\varphi\right)\right),\label{eq:36}
\end{equation}
and therefore the scale factor grows by a huge factor 
\begin{equation}
\frac{a_{f}}{a_{i}}\simeq\exp\left(\frac{1}{\sqrt{6}m}\right)\gg1,\label{eq:37}
\end{equation}
for $m\ll1$ during inflation.

Let us now turn to the self-reproduction condition in this model.
During the typical Hubble time $t_{H}\simeq H^{-1}$, the classical
background field decreases by 
\begin{equation}
\delta\varphi_{cl}\simeq\dot{\varphi}t_{H}\simeq\frac{\dot{\varphi}}{H}\sim-1,\label{eq:38}
\end{equation}
i.e., always by the order of the Planck value (cf. (\ref{eq:5})).
Therefore, taking into account the super-imposed positive quantum
fluctuation of amplitude of order $H$, the total change in the value
of the scalar field at the Hubble scale is equal 
\begin{equation}
\Delta\varphi\simeq H-1<0,\label{eq:39}
\end{equation}
i.e., negative and the field always decreases at sub-Planckian scales.
Thus, the condition for self-reproduction is never satisfied, and
we avoid an eternally inflating universe.

To conclude this subsection, let us derive the conditions that a general
potential $V\left(\varphi\right)$ must satisfy to produce an inflationary
stage of the type considered above. In order to find an analog of
the slow-roll conditions in our case, we assume that the second derivative
of the scalar field in equation (\ref{eq:25}) can be neglected. The
approximate solution of this equation is then 
\begin{equation}
\dot{\varphi}\approx-\frac{V^{\prime}}{\kappa_{0}}.\label{eq:40}
\end{equation}
Using this solution and taking into account (\ref{eq:26}), we find
that this approximation is justified, i.e., $\left\vert \ddot{\varphi}\right\vert \ll\left\vert \kappa\dot{\varphi}\right\vert $
only if 
\begin{equation}
\frac{V^{\prime\prime}}{V^{\prime}}\ll1.\label{eq:41}
\end{equation}
The relative rate of the change of the Hubble scale is equal to 
\begin{equation}
\frac{\left\vert \dot{\kappa}\right\vert }{\kappa^{2}}\simeq\frac{\left\vert \ddot{\varphi}\right\vert }{\left\vert \kappa\dot{\varphi}\right\vert }\simeq\frac{V^{\prime\prime}}{V^{\prime}}\ll1,\label{eq:42}
\end{equation}
i.e., we have an exponential expansion if the inequality (\ref{eq:41})
is satisfied. For any power-law potential, this inequality holds at
$\varphi>1$ and the self-reproduction condition is satisfied only
when the curvature becomes larger than the Planck curvature.

\subsection{Case B}

It follows from the CMB observations that the potential $V$ in the
region responsible for observable scales tends to the flat potential
and the ratio of the tensor to scalar perturbations must be small.
As we will see in the next section, this ratio is always of order
one in the models considered above and are therefore not realistic
and must be modified. There are many potentials that satisfy the required
conditions and are consistent with observations and do not exhibit
self-reproduction. For illustration, in this subsection we will consider
one of the simplest potentials of this kind. Namely, we take the potential
\begin{equation}
V\left(\varphi\right)=\frac{1}{2}\frac{m^{2}\varphi^{2}}{\left(1+\varphi^{2}\right)}\left(1+m\varphi^{4}\right),\label{eq:43}
\end{equation}
and the function corresponding to the mimetic interactions 
\begin{equation}
C\left(\kappa\right)=1+\frac{\kappa}{m},\label{eq:44}
\end{equation}
and assume that $m\ll1.$ For small $\varphi<1$ this potential describes
a massive scalar field, while for $\varphi>1$ it can be well approximated
as 
\begin{equation}
V\simeq\frac{1}{2}m^{2}\left(1-\frac{1}{\varphi^{2}}+m\varphi^{4}\right).\label{eq:45}
\end{equation}
In this approximation, equations (\ref{eq:20})-(\ref{eq:21}) become
\begin{align}
 & \kappa^{2}=\frac{3}{2}m^{2}\left(1-\frac{1}{\varphi^{2}}+m\varphi^{4}\right)+\frac{3}{2}\dot{\varphi}^{2},\label{eq:46}\\
 & \ddot{\varphi}+\kappa\dot{\varphi}+\left(1+\frac{\kappa}{m}\right)\left(\frac{m^{2}}{\varphi^{3}}+2m^{3}\varphi^{3}\right)=0.\label{eq:47}
\end{align}
First, we determine the region where the potential term in (\ref{eq:46})
is dominant compared to the kinetic term. We assume that the slow-roll
approximation is valid (this has to be checked afterwards) and that
the second derivative of the scalar field in equation (\ref{eq:47})
can be neglected, hence 
\begin{equation}
\dot{\varphi}\simeq-\left(\frac{1}{\kappa}+\frac{1}{m}\right)\left(\frac{m^{2}}{\varphi^{3}}+2m^{3}\varphi^{3}\right)\simeq-\left(\frac{m}{\varphi^{3}}+2m^{2}\varphi^{3}\right),\label{eq:48}
\end{equation}
because $\kappa>m$ at $\varphi>1$ as can be seen from (\ref{eq:46}).
Comparing $\dot{\varphi}^{2}$ with $V$ we find that for 
\begin{equation}
1<\varphi<m^{-\frac{1}{2}},\label{eq:49}
\end{equation}
indeed $\dot{\varphi}^{2}<V$ holds and inflation is dominantly driven
by the potential. One can easily check that in the whole range (\ref{eq:49})
the condition for self-reproduction is not fulfilled, i.e. the decrease
of the classical background field during the typical Hubble time ,
$\delta\varphi_{cl}\simeq\dot{\varphi}t_{H}$, is always larger than
the amplitude of the quantum fluctuations in the Hubble scale, so
that the scalar field decreasing in total. The energy density at $\varphi\simeq m^{-1/2}$
is about $\varepsilon\simeq m.$ At $\varphi>m^{-1/2}$ the kinetic
term in (\ref{eq:46}) becomes dominant, and the slow-roll condition
for the scalar field is still satisfied, so that, 
\begin{equation}
\dot{\varphi}\simeq-2m^{2}\varphi^{3},\label{eq:50}
\end{equation}
while 
\begin{equation}
\kappa^{2}\simeq\frac{3}{2}\dot{\varphi}^{2},\label{eq:51}
\end{equation}
and we obtain the inflationary stage of the type described in subsection
A. Up to the Planck scale, which is reached at $\varphi\simeq m^{-2/3}$,
self-reproduction does not occur. Therefore. at the Planck scale,
we can begin inflation, which simultaneously solves the problem of
fine-tuning and avoids an eternally self-reproducing universe. As
we will show in the next section, the considered model satisfies all
constraints imposed by observations and does not lead to problems
related to an eternal universe. One must keep in mind that this is
the simplest example of such models and one can easily construct many
of them, all in agreement with current CMB observations.

\section{Perturbations}

Since for first-order in perturbations the spatial components of $\delta T_{k}^{i}$
vanish for $i\neq k$, the perturbed metric can be written in conformal
Newtonian gauge as follows (see, for example, \cite{Mukhanovbook})
\begin{equation}
ds^{2}=\left(1+2\Phi\right)dt^{2}-a^{2}\left(t\right)\left[\left(1-2\Phi\right)\delta_{ik}dx^{i}dx^{k}-h_{ik}^{(t)}dx^{i}dx^{k}\right],\label{eq:52}
\end{equation}
where $\Phi$ is the gravitational potential for the scalar perturbations
and the transverse, traceless part of the metric $h_{ik}^{(t)}$ describes
gravitational waves. The equation for gravitational waves remain the
same as in general relativity, so we do not need consider them further
here and refer the reader to, for example, reference \cite{Mukhanovbook}.
However, the consideration of the scalar perturbations in our mimetic
inflation is substantially modified. For the linear order in perturbations,
the constraint equation (\ref{eq:7}) gives 
\begin{equation}
\dot{\delta\phi}=\Phi,\label{eq:53}
\end{equation}
where $\delta\phi$ is the perturbation of the mimetic field. First,
we note that 
\begin{equation}
\delta T_{i}^{0}=\left(2\lambda+\left(C^{\prime}V\right)\dot{}\right)\delta\phi_{,i}+\delta\left(C^{\prime}V\right)_{,i}+\dot{\varphi}\delta\varphi_{,i}.\label{eq:54}
\end{equation}
The first term vanishes here in the absence of mimetic dust (see (\ref{eq:19})).
Noting that 
\begin{align}
 & \delta\left(C^{\prime}V\right)=C^{\prime\prime}V\delta\kappa+C^{\prime}V^{\prime}\delta\varphi\nonumber \\
 & =-3C^{\prime\prime}V\left(\dot{\Phi}+H\Phi+\frac{1}{3a^{2}}\Delta\delta\phi\right)+C^{\prime}V^{\prime}\delta\varphi,\label{eq:55}
\end{align}
where $H=\kappa/3$ is the Hubble constant and we used the relation
(\ref{eq:53}) to express the derivatives of the mimetic field perturbations
in terms of the gravitational potential, we obtain from the $0-i$
Einstein equations 
\begin{align}
 & \dot{\Phi}+H\Phi=\frac{1}{2\left(1+\frac{3}{2}C^{\prime\prime}V\right)}\left[\left(\dot{\varphi}+C^{\prime}V^{\prime}\right)\delta\varphi-\frac{C^{\prime\prime}V}{a^{2}}\Delta\delta\phi\right]\nonumber \\
 & =-\frac{\dot{H}}{\dot{\varphi}}\delta\varphi-\frac{C^{\prime\prime}V}{\left(2+3C^{\prime\prime}V\right)a^{2}}\Delta\delta\phi,\label{eq:55a}
\end{align}
where we also used equation (\ref{eq:23}). The linearized equation
(\ref{eq:15}) for the inflaton perturbations is 
\begin{equation}
\ddot{\delta\varphi}+3H\dot{\delta\varphi}-\frac{1}{a^{2}}\Delta\delta\varphi-4\dot{\varphi}\dot{\Phi}-2\left(\ddot{\varphi}+3H\dot{\varphi}\right)\Phi+\delta\left(CV^{\prime}\right)=0.\label{eq:57}
\end{equation}
Similar to (\ref{eq:55}), we find 
\begin{equation}
\delta\left(CV^{\prime}\right)=-3C^{\prime}V^{\prime}\left(\dot{\Phi}+H\Phi+\frac{1}{3a^{2}}\Delta\delta\phi\right)+CV^{\prime\prime}\delta\varphi.\label{eq:58}
\end{equation}
Taking the time derivative of (\ref{eq:21}) we can express $CV^{\prime\prime}$
as follows 
\begin{equation}
CV^{\prime\prime}=-\frac{1}{\dot{\varphi}}\left(\dddot{\varphi}+3H\ddot{\varphi}+3\dot{H}\dot{\varphi}+3C^{\prime}V^{\prime}\dot{H}\right)\label{eq:59}
\end{equation}
Substituting this into (\ref{eq:57}), after replacing $\dot{\Phi}+H\Phi$
in the obtained equation with (\ref{eq:55a}) we we can rewrite the
equation (\ref{eq:57}) as follows 
\begin{align}
 & \ddot{\delta\varphi}+3H\dot{\delta\varphi}-\frac{1}{a^{2}}\Delta\left(\delta\varphi+\frac{2C^{\prime}V^{\prime}-4\dot{\varphi}C^{\prime\prime}V}{2+3C^{\prime\prime}V}\delta\phi\right)\nonumber \\
 & -\left(\frac{\dddot{\varphi}}{\dot{\varphi}}+3H\frac{\ddot{\varphi}}{\dot{\varphi}}-\dot{H}\right)\delta\varphi-2\left(\ddot{\varphi}+H\dot{\varphi}\right)\Phi=0.\label{eq:60}
\end{align}
As can be seen, $0-0$ and $i-i$ Einstein equations do not provide
any additional useful information. Substituting the expression for
$\delta\varphi$ in terms of $\Phi$ and $\delta\phi$ from (\ref{eq:55a})
into $i-i$ Einstein equation, we find that it is identically satisfied.
Using the $0-0$ Einstein equation, we can determine $\delta\lambda$
, which we are not interested in. If we take the time derivative of
this equation and use the linearized version of the $\phi$-equation
(\ref{eq:14}), we come back to (\ref{eq:60}). Thus, the three equations
(\ref{eq:53}), (\ref{eq:55a}) and (\ref{eq:60}) are sufficient
to completely determine the unknown functions $\delta\varphi,\Phi$
and $\delta\phi$. Let us consider the plane wave with co-moving wave-number
$k=\left\vert \overrightarrow{k}\right\vert $, i.e., 
\begin{equation}
\delta\varphi,\Phi,\delta\phi\propto\exp\left(i\overrightarrow{k}.\overrightarrow{x}\right).\label{eq:61}
\end{equation}
In this case, the behavior of the perturbations depends drastically
on whether the physical wavelength $\lambda_{ph}\simeq a/k$ is much
smaller or much larger compared to the Hubble scale $H^{-1}.$ For
the short-wavelength perturbations with $k\gg Ha$, we simplify (\ref{eq:60})
to the equation 
\begin{equation}
\ddot{\delta\varphi_{k}}+3H\dot{\delta\varphi_{k}}+\frac{k^{2}}{a^{2}}\delta\varphi_{k}\simeq0,\label{eq:62}
\end{equation}
which has a simple solution 
\begin{equation}
\delta\varphi_{k}\simeq\frac{A_{k}}{a}\exp\left(\pm ik\int\frac{dt}{a}\right),\label{eq:63}
\end{equation}
where $A_{k}$ is the constant of integration. We now need to verify
that the skipped terms for $k\gg Ha$ are really negligible. Since
$\Phi$ and $\delta\phi$ also oscillate, we can estimate their derivatives
at the leading order as follows: 
\begin{equation}
\dot{\Phi}\sim\frac{k}{a}\Phi,\;\dot{\delta\phi}\sim\frac{k}{a}\delta\phi,\label{eq:64}
\end{equation}
Using these estimates in equations (\ref{eq:53}) and (\ref{eq:55a})
we find that 
\begin{equation}
\frac{2C^{\prime}V^{\prime}-4\dot{\varphi}C^{\prime\prime}V}{2+3C^{\prime\prime}V}\delta\phi\sim\frac{\dot{H}}{\left(k/a\right)^{2}}\delta\varphi,\label{eq:65}
\end{equation}
and therefore the terms containing perturbations of the mimetic field
$\delta\phi$ and, and Laplacian $\Delta$ in (\ref{eq:60}) could
be neglected compared to $\delta\varphi$ for $k/a\gg H$. The other
skipped terms in equation (\ref{eq:60}) are also suppressed by at
least a factor $Ha/k\ll1$ compared to the terms retained in (\ref{eq:62})
by at least a factor $Ha/k\ll1$. After the inhomogeneity crosses
the Hubble scale at the time $t_{k}$ determined from the equation
$k\sim H_{k}a_{k}$, the spatial derivatives terms in (\ref{eq:60})
decay as $1/a^{2}$ and the Laplacian in (\ref{eq:55a}) and (\ref{eq:60})
can be neglected, so the resulting equations become 
\begin{align}
 & \dot{\Phi}+H\Phi\simeq-\frac{\dot{H}}{\dot{\varphi}}\delta\varphi\nonumber \\
 & \ddot{\delta\varphi}+3H\dot{\delta\varphi}-\left(\frac{\dddot{\varphi}}{\dot{\varphi}}+3H\frac{\ddot{\varphi}}{\dot{\varphi}}-\dot{H}\right)\delta\varphi-2\left(\ddot{\varphi}+H\dot{\varphi}\right)\Phi\simeq0,\label{eq:67}
\end{align}
for $k\ll Ha$. The exact solutions of these equations, as can be
easily proved by direct substitution, are 
\begin{equation}
\delta\varphi=A\frac{\dot{\varphi}}{a}\int adt,\quad\Phi=A\frac{d}{dt}\left(\frac{1}{a}\int adt\right),\label{eq:68}
\end{equation}
where $A$ is an integration constant, and as follows from (\ref{eq:53}),
the perturbation of the mimetic field is 
\begin{equation}
\delta\phi=A\frac{1}{a}\int adt.\label{eq:69}
\end{equation}
Note that the above solutions for any arbitrary $C(\kappa)$ and $V(\varphi)$
are valid not only during inflation but also after the end of inflation.

\section{Spectrum of inhomogeneities}

Now we can calculate the spectrum of inhomogeneities arising from
initial quantum fluctuations during mimetic inflation. First, note
that the quantum minimal fluctuations are well defined only on scales
smaller than the Hubble scale, on which space-time can be well approximated
by the flat Minkowski metric. Moreover, the initial spectrum of perturbations
on these scales need not be fine-tuned at the outset. Indeed, it may
initially contain particles, but these particles and existing inhomogeneities
must not destroy the quasi-exponential expansion from the very beginning.
Later, these inhomogeneities, present at sub-Hubble scales, are stretched
by expansion to very large unobservable scales and become completely
irrelevant. Therefore, there is \textit{no fine-tuning problem at
all} even if the inflating region is largely inhomogeneous, under
the only condition that this initial inhomogeneity does not prevent
inflationary expansion from the beginning.

The other two problems, namely the Trans-Planckian problem and the
choice of the Bunch-Davies vacuum, are widely discussed in the literature.
\footnote{We do not give here references because there is huge literature with
erroneous statements on this topic and the reader should better identify
the relevant publications with an appropriate search engine.} After a relatively short time of inflation, the inhomogeneities on
scales $\lambda_{ph}<H^{-1}$ are the quantum fluctuations. Thus,
contrary to many statements in the literature, one does not need to
postulate a Bunch-Davies or even a Minkowski vacuum at these scales
at the beginning of inflation. As emphasized above, the only requirement
needed is that the exponential expansion should not be terminated
at the very beginning, since as time goes on the exponential expansion
makes these initial inhomogeneities less and less relevant to the
evolution of the universe at observable scales. The Trans-Planckian
problem is formulated very similarly for Hawking radiation and inflationary
perturbations. As can be seen, it is an artificial problem due to
the calculational peculiarities of the standard derivations of these
effects. In fact, after cleaning up the initial inhomogeneities on
sub-Hubble scales, only the inevitable quantum fluctuations on scales
$\lambda_{ph}<H^{-1}$ remain as a result of the expansion. In the
static coordinate system, which can always be introduced within the
Hubble scale, this is simply well-known Minkowski vacuum. It is convenient
to describe this vacuum in an expanding coordinate system, such as
done in equation (\ref{eq:63}). The only purpose of this description
has a technical reason, because it allows us to relate the sub-Hubble
scales with the scales exceeding the Hubble scale, where the static
coordinate system does not exist. For the scalar field, the spectrum
of fluctuations, i.e., the dependence of the typical amplitude on
the physical scale $\lambda_{ph}\simeq a/k$ for $\lambda_{ph}<H^{-1}$
remains invariant and does not change during the ``evolution''\ described
by equation (\ref{eq:62}). In particular, for the massive scalar
field $\delta\varphi\equiv\sqrt{\delta\varphi_{k}^{2}k^{3}}\simeq1/\lambda_{ph}$
and it is equal to $\delta\varphi\simeq H$ on the Hubble scale. Following
the standard calculations in expanding coordinate system, one gets
the impression that the perturbations from a certain physical scale
are removed by expansion and replaced by the perturbations whose physical
wavelength was originally smaller than the Planck scale. This preserves
the invariant vacuum spectrum (see, e.g. \cite{Mukhanovbook}). It
is so called Trans-Planckian problem. However, this is just a simple
technical trick to simplify the calculations. Instead of thinking
this way, one could simply say that instead of the perturbations ``stolen
by expansion''\ from physical scales $\lambda_{ph}<H^{-1}$, the
uncertainty relation simply generates the new perturbation with the
required amplitude, so that in the static coordinate system the vacuum
spectrum remains invariant at sub-Hubble scales. In this case we do
not even need to talk about the fact that the perturbations had the
Trans-Planckian scales and were later stretched by expansion into
the corresponding physical scales. The situation here is very similar
to that when we describe the Minkowski vacuum in expanding Milne coordinates.
Therefore, the Trans-Panckian problem is rather a technical artifact
of derivation than a real physical problem. From the cosmic inflation
point of view, the most important fact is that there are always unavoidable
quantum fluctuations with amplitude $\delta\varphi\simeq H$ on the
scales $\lambda_{ph}\simeq H^{-1}$, which are used as initial conditions
for the perturbation with co-moving wave-number $k$ and at the time
$t_{k}$ satisfying $k\simeq a\left(t_{k}\right)H\left(t_{k}\right)$,
when the physical wavelength of this perturbation is of the order
of the Hubble scale. After that, the perturbation can only be described
by expanding coordinates, since the static coordinate system no longer
exists on scales exceeding the Hubble scale. For $t>t_{k}$, the scale
factor grows exponentially, the Hubble constant does not change significantly,
and the perturbation satisfies the condition $k<Ha,$ and is therefore
described by the solution (\ref{eq:68}).

During inflation, $\dot{H}\ll H^{2}$, and therefore the integral
in (\ref{eq:68}) can be well approximated as 
\begin{equation}
\frac{1}{a}\int adt=\frac{1}{a}\int\frac{da}{H}=\frac{1}{H}\left(1+\frac{\dot{H}}{H^{2}}+...\right)+\frac{D}{a}\simeq\frac{1}{H},\label{eq:70}
\end{equation}
where we also neglected the decaying mode with the integration constant
$D$. Considering the perturbation with the co-moving wave-number
$k$ and taking into account that the typical amplitude of quantum
fluctuations is $\sqrt{\delta\varphi_{k}^{2}k^{3}}\simeq H_{k=Ha}$
at time $t_{k}$, it follows from the first equation in (\ref{eq:68})
that the integration constant is equal to 
\begin{equation}
A\simeq\left(\frac{H^{2}}{\left\vert \dot{\varphi}\right\vert }\right)_{k=Ha},\label{eq:71}
\end{equation}
where the subscript means that the corresponding quantity in the parenthesis
must be evaluated at the time when the perturbation is stretched by
inflationary expansion to the physical scale $\lambda_{ph}\simeq H^{-1}.$
In the further estimates, we will omit all coefficients of order one
(see, e.g., \cite{Mukhanovbook} for rigorous definitions), since
they affect only the total amplitude of the spectrum, which is a free
parameter of the theory determined by observations. From this and
from (\ref{eq:68}) it follows that 
\begin{equation}
\Phi\simeq\left(\frac{H^{2}}{\left\vert \dot{\varphi}\right\vert }\right)_{k=Ha}\frac{d}{dt}\left(\frac{1}{a}\int adt\right).\label{eq:72}
\end{equation}
After the end of the inflation, the last time dependent term in this
equation becomes a constant of order one, and therefore, for the perturbations
that left the Hubble scale during inflation, the spectrum of gravitational
potential $\delta_{\Phi}\equiv\sqrt{\Phi_{k}^{2}k^{3}}$ is given
by 
\begin{equation}
\delta_{\Phi}\simeq\left(\frac{H^{2}}{\left\vert \dot{\varphi}\right\vert }\right)_{k=Ha}.\label{eq:73}
\end{equation}
The consideration of gravitational waves in mimetic inflation is exactly
the same as in ordinary inflation, and therefore we simply quote the
final result for them (see, e.g., \cite{Mukhanovbook}) 
\begin{equation}
\delta_{h}\simeq H_{k=Ha}.\label{eq:74}
\end{equation}
Now we will apply the obtained results to calculate the spectrum of
perturbations in the two particular models considered in the previous
section.

\subsection{Case A}

For the case where $C=\kappa/\kappa_{0}$ it follows from (\ref{eq:25})
and (\ref{eq:40}) that during inflation 
\begin{equation}
\left\vert \dot{\varphi}\right\vert \simeq H\simeq\frac{V^{\prime}}{\kappa_{0}},\label{eq:75}
\end{equation}
and therefore 
\begin{equation}
\delta_{\Phi}\simeq H_{k=Ha}\simeq\left(\frac{V^{\prime}}{\kappa_{0}}\right)_{k=Ha}.\label{eq:76}
\end{equation}
In this model, the amplitudes of the scalar perturbations and gravitational
waves (see Eq. (\ref{eq:74})) are the same up to a numerical factor
of order one at all scales. In those scales that were of order $H^{-1}$
at the beginning of inflation, i.e., at Planck scales the amplitude
of both scalar perturbations and gravitational waves is of order one
and it decreases to the value $V^{\prime}\left(\varphi\simeq1\right)/\kappa_{0}$
for scales that crossed the Hubble scale at the end of inflation.
Note that the amplitude of scalar perturbations is never larger than
unity and is consistent with the absence of self-reproduction. As
can be seen from observations, the amplitude of the gravitational
waves at the observable scales is much smaller than the amplitude
of the scalar perturbations and therefore this model is definitely
ruled out.

\subsection{Case B }

\begin{figure}[ht]
\includegraphics[width= 12.6 cm]{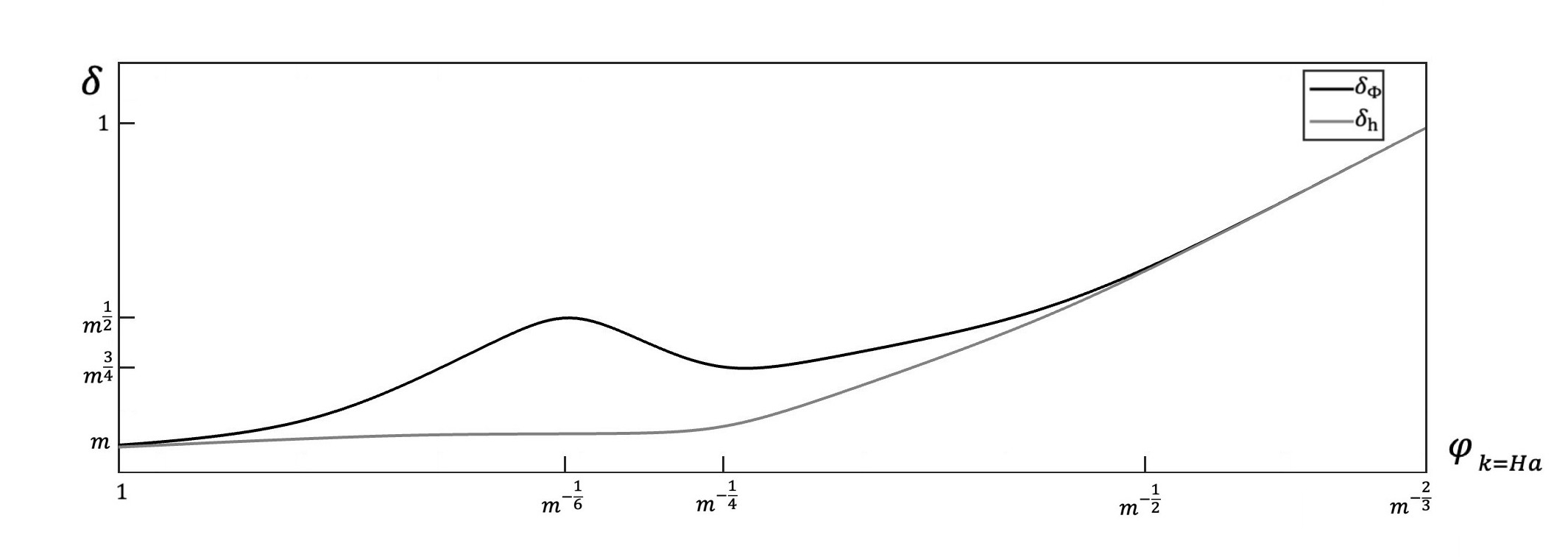}\caption{Amplitudes of the spectrum for scalar perturbations and gravitational waves}
\centering
\label{Fig 1}
\end{figure}

Now we turn to a more realistic model where $V$ and $C$ are given
in (\ref{eq:23}) and (\ref{eq:24}), respectively. First, we consider
\begin{equation}
1<\varphi<m^{-1/2},\label{eq:77}
\end{equation}
where the potential term dominates compared to the kinetic term in
equation (\ref{eq:46}). In this region, the consideration of perturbations
is similar to the standard potential-dominated inflation, except for
the expression (\ref{eq:49}) for $\dot{\varphi}$, which is different
from the standard expression for $\kappa\gg m$. Substituting $H^{2}$
and $\dot{\varphi}$ from (\ref{eq:45}) and (\ref{eq:48}) respectively,
into (\ref{eq:73}), we get 
\begin{equation}
\delta_{\Phi}\simeq m\left(\varphi^{3}\frac{1+m\varphi^{4}}{1+2m\varphi^{6}}\right)_{k=Ha}.\label{eq:78}
\end{equation}
To express $\varphi_{k=Ha}$ for a given co-moving wave-number $k$,
we first introduce the number of e-folds $N$ before the end of inflation
\begin{equation}
a\simeq a_{f}e^{-N},\label{eq:79}
\end{equation}
where $a_{f}$ is the scale factor at the end of inflation. Then the
perturbation with a given $k$ starts to exceed the Hubble scale,
i.e. the condition $k=Ha$ is satisfied, at 
\begin{equation}
N_{k}\simeq\ln\left(\frac{Ha_{f}}{k}\right)\label{eq:80}
\end{equation}
e-folds before inflation ends. For the scales corresponding to the
range covered by the CMB observations $50<N_{k}<60$, the exact values
of which depends, e.g., on the details of the reheating immediately
after the end of inflation.\footnote{The exact value of $N$ depends on the unknown physics beyond the
Standard Model of particle physics, which will most likely never be
clarified, since we are dealing with energy scales that can never
be reached with accelerators. For this reason, further improvement
of the already achieved accuracy in the detemination, e.g., of the
spectral index of perturbations will not give much useful information
for the selection of the right ``fundamental physics''\ scenario
for inflation.} To obtain the relation between $N_{k}$ and $\varphi_{k=Ha}$, we
note that 
\begin{equation}
H=-\frac{dN}{d\varphi}\dot{\varphi},\label{eq:81}
\end{equation}
from where it follows that 
\begin{equation}
N=-\int\frac{H}{\dot{\varphi}}d\varphi\simeq\int\frac{\varphi^{3}\left(1+m\varphi^{4}\right)^{1/2}}{1+2m\varphi^{6}}d\varphi.\label{eq:82}
\end{equation}
In the interval $1<\varphi<m^{-1/6}$ we find that $N_{k}\simeq\varphi_{k=Ha}^{4}$
and from (\ref{eq:78}) it follows that 
\begin{equation}
\delta_{\Phi}\simeq m\varphi_{k=Ha}^{3}\simeq mN_{k}^{3/4},\label{eq:83}
\end{equation}
and the spectral index is 
\begin{equation}
n_{s}-1\equiv\frac{d\ln\delta_{\Phi}^{2}}{d\ln k}=-\frac{d\ln\delta_{\Phi}^{2}}{dN_{k}}=-\frac{3}{2N_{k}}.\label{eq:84}
\end{equation}
For $N_{k}=50$ we have $n_{s}=0.97$ in agreement with the observations.
To obtain the correct amplitude in the observable scales, we must
take $m\simeq10^{-6}$ and the range of scales where the formula (\ref{eq:84})
is valid corresponds to $1<N_{k}<10^{4}$, well beyond the cosmological
horizon today. The tensor-to-scalar ratio 
\begin{equation}
r\equiv\frac{\delta_{h}^{2}}{\delta_{\Phi}^{2}}\propto\frac{1}{N^{3/2}},\label{eq:85}
\end{equation}
is less suppressed in our case than in $R^{2}$ and Higgs inflationary
scenarios \cite{StarHigs}, where $r\propto1/N^{2}$, and still in
agreement with the present observational upper bound on $r.$ The
model we consider can be easily modified by the choice of the potential
$V$, so that it will fully agree with \cite{StarHigs} at observable
scales. On scales corresponding to $m^{-1/6}<\varphi_{k=Ha}<m^{-1/4}$,
the perturbation amplitude decreases as $\delta_{\Phi}\simeq\varphi_{k=Ha}^{-3}$
as the scale increases, changing its value from $m^{1/2}$ to $m^{3/4}$.
At even larger scales corresponding to $m^{-1/4}<\varphi_{k=Ha}<m^{-1/2}$,
the amplitude starts to grow again as $\delta_{\Phi}\simeq m\varphi_{k=Ha}$
and becomes $\delta_{\Phi}\simeq\delta_{h}\simeq m^{1/2}$ at $\varphi\simeq m^{-1/2}$.
For $\varphi>m^{-1/2}$, the kinetic term dominates compared to the
potential term, and for these values of $\varphi$, inflation proceeds
as described in subsection A of section 3. For scales that have left
the Hubble scale at this stage, i.e., at the beginning of inflation,
$\delta_{\Phi}\simeq\delta_{h}\simeq m^{2}\varphi_{k=Ha}^{3}$ the
amplitude of both gravitational waves and scalar perturbations is
the same. It becomes of order one only at Planck densities at $\varphi\simeq m^{-2/3}$.
The amplitudes of the spectrum for the scalar perturbations and for
the gravitational waves as a function of scale corresponding to $\varphi_{k=Ha}$,
are shown in Figure \ref{Fig 1}. Note that below the Planck scale
they are never larger than one, which is consistent with the condition
of no self-reproduction. 


\section{Conclusions}

We have shown how self-reproduction and eternal inflation could be
easily avoided if we couple the inflaton potential to the mimetic
field. We would like to emphasize again that this does not mean at
all that we have added an additional scalar field, but that we have
modified Einstein gravity at high curvatures. In fact, the only additional
degree of freedom for the mimetic field is a ``dust''\ that becomes
negligible soon after inflation begins, but can be generated later
to account for the dark matter component in our universe. Note that
self-reproduction can be avoided if we start inflation at Planck curvature
(to avoid any kind of fine-tuning), if the function $C\left(\kappa\right)$
introduced in (\ref{eq:11}) is a linear function of $\kappa$ at
large curvature. Only in this case the main contribution to the energy
density of the inflaton field comes entirely from the kinetic energy
and as a result the amplitudes of the generated gravitational waves
and scalar perturbations become comparable on the corresponding scales.

At smaller curvatures, this function is rather arbitrary and this
opens the possibility of constructing a whole class of inflationary
scenarios. Moreover, the Lagrangian (\ref{eq:11}) can be further
generalized by adding an extra potential $\tilde{V}\left(\varphi\right)$,
which is different from $V\left(\varphi\right)$, and even by coupling
the scalar field directly to the first derivative of the mimetic field,
such as, for example $g^{\mu\nu}E\left(\kappa\right)\partial_{\mu}F\left(\varphi\right)\partial_{\nu}\phi$.
This greatly expands the scope for constructing simple inflationary
scenarios. We will leave the detailed analysis of such models to further
investigation.

\textbf{\large{}{}{}{}{}{Acknowledgments}}{\large\par}

A.H.C would like to thank Ludwig Maximillians University, Munich, for
hospitality where part of this work is done. The work of A.H.C. is
supported in part by the National Science Foundation Grant No. Phys-2207663.
The work of M.K. and V.M. is supported by the Deutsche Forschungsgemeinschaft
(DFG, German Research Foundation) under Germany's Excellence Strategy
-- EXC-2111 -- 390814868.

\end{document}